\documentclass[a4paper,12pt]{article}

\usepackage[pdftex]{graphicx}
\usepackage{authblk}

\title{Complete Decentralization of Distributed Data Storages Based on Blockchain Technology}

\author{A.P.Demichev\thanks{demichev@theory.sinp.msu.ru}, A.P.Kryukov\thanks{kryukov@theory.sinp.msu.ru}}

\affil{\small\textit{Skobeltsyn Institute of Nuclear Physics,}\\ \textit{Lomonosov Moscow State University, Moscow, Russia}}

\date{}

\begin{document}

\maketitle
\begin{abstract}
The work presents a solution for completely decentralized data management systems in geographically distributed environments with administratively unrelated or loosely related user groups and in conditions of partial or complete lack of trust between them. The solution is based on the integration of blockchain technology, smart contracts and provenance metadata driven data management. Architecture, operation principles and algorithms developed provides fault-tolerant, safe and reliable management of provenance metadata, control of operations with data files, as well as resource access management in collaborative distributed computing systems. The latter refer to distributed systems formed by combining into a single pool of computer resources of various organizations (institutions) to work together in the framework of some project.
\end{abstract}


\section{Introduction}
The infrastructure of most distributed computer systems (DCS) contains system-wide centralized services (see, \textit{e.g.} \cite{1,2,3}). In particular, such services that control the operation of DCS in general are: data management services, system monitoring services, services providing access to metadata and searching for data by the metadata filters, services  for automatic renewal of proxy certificates, etc. Thus, although local data storages and processing resources are distributed in the DCS environment, they are integrated into a single pool using the infrastructure based on centralized services. The latter are potentially points of failure, malicious intrusion, seizing control of DCS and/or the bottlenecks for the system. In addition, users are forced to trust central services administrators in matters of access control, management and use of data. Therefore, an important problem is to develop fully decentralized methods for managing data and access rights to them in DCS environment.

One approach to the decentralization is based on the use of peer-to-peer (P2P) networks \cite{4}. However, a significant problem of this approach when using it for a specific and time-limited project is to provide a stable pool of peers, that is, storage resource providers, especially at the initial stage of development of such a network. In other words, before such a P2P-based storage can work stably, it will require significant technical, organizational and time costs from the organizers in the absence of a guaranteed result, that is, an operating network with sufficient storage capacity.

Another approach, considered in this paper, is applicable to collaborative distributed systems. We use the term collaborative distributed computing system to refer to distributed systems formed by combining into a single pool of computer resources of various organizations (institutions) to work together within a project and, possibly, in conditions of complete or partial lack of trust between user groups. In this case, it is very fruitful to use blockchain technologies for data management based on consensus between the parties involved on the content and order for data operations. 

In this paper, only the basic principles of the approach are presented; all the details of our approach to data storing and processing in a collaborative distributed environment as well as its implementation can be found in the work \cite{9}.

\section{Decentralized collaborative distributed systems}

In the case of large-scale projects related to data handling and involving participants from different administrative institutions the problem arises of combining participants’ resources, including storages, data in them and processing services, into a unified distributed computing system (DCS), as well as ensuring the implementation of mutual data access policies of the parties involved. For brevity, we will refer to systems of this type as collaborative distributed computing systems (CDCS). Operation of such system requires methods for decentralized management of data and access rights to them in a dynamically changing environment, ensuring consensus among the parties on the content and order of data operations and ensuring reliable, immutable recording of proven operations history, that is, provenance metadata (PMD). The latter are necessary for data storage and usage consistency, as well as for consideration and resolving possible conflicts among project participants or with the storage/service providers. In other words, it is necessary to provide tools to support the implementation of business processes of data storing and processing in a distributed environment and in the presence of administratively unrelated or loosely related organizations involved in joint projects, or simply sharing data under certain conditions. In the works \cite{3,5} we proposed a new approach to the construction a management system for provenance metadata and data access rights in a distributed environment, based on the integration of blockchain technology, smart contracts and provenance metadata driven data management. We also suggested a new method for delegation of rights from a user or service to another service within distributed computing systems. The implementation of the proposed approach, entitled ProvHL, is based on the permissioned blockchains and on the Hyperledger Fabric blockchain platform  (HLF;  www.hyperledger.org) \cite{6}. 

This platform manages the permissioned type of blockchains intended for a limited number of parties in the business process. In general, two approaches are possible: (1) data management systems (DMS) of local storages manage data and use blockchain as a distributed ledger (data driven data management); (2) first, metadata are written to the blockchain, and DMSs refer to the blockchain and performs the transactions recorded there (metadata driven data management). In the first case, the functionality of the blockchain system is very limited,  it only provides resistance to malicious attempts to modify the history of data in distributed storage. The HLF platform enables one to implement the second approach, which in addition to simply maintaining the distributed ledger allows us to control operations with data in DCS and solve the problem of decentralized data management. Basic operations comprise of a set of atomic transactions and can be of the following types: new file upload;  file download; file copy within a local storage; file deletion; file copy to another local storage; file transfer to another local storage. In addition to the operations related only to file storage, in many cases, users need to perform a large number of much the same tasks of data processing. Often such tasks are identical in form, differ only in the values of the input data (task parameters) and can be performed by trusted pre-installed specialized web services. Within the framework of the ProvHL environment, we can consider data processing by such web services as another type of file operations. Each active transaction, and therefore operation, corresponds to an update of some file state attributes, that is PMD. 

To describe the business process within the framework of HLF platform, a number of concepts are used, the main ones are assets, participants, transactions and events. For their description in the framework of the HLF platform a special modeling language is used. The most important assets in our case are the data files in a distributed storage. Their properties (attributes) are the provenance metadata. Other types of the assets are directories, operations and groups. Participants are members of the business network. They can own assets and make transaction requests. The transaction is the mechanism of interaction of participants with assets. Event messages can be sent by transaction processors to inform external software components of changes in the blockchain; applications can subscribe to receive event information via the HLF API.

The definition of the operations as assets allows us to keep track of the operation own history (the states of the operations are 'started', 'pending', 'completed', 'error', etc.). Also this improves the level of correspondence between the history recorded in the blockchain and the real history of the data in a DCS as well as makes the mechanism of the delegation of rights in a distributed environment very natural and flexible \cite{7}. Each operation with data comprises of at least two types of transactions: one corresponds to the client request and the second corresponds to the server response.  Accordingly, at least two records are made in the blockchain. For example, when the upload operation is performed, the transaction on file creation, initiated by the client request, is first recorded in the blockchain with the "temporary" label. And only after the actual uploading the file in the storage, data management server initiates a transaction removing the label "temporary" and turning the uploaded file into a fully valid asset. This makes the level of correspondence between the history recorded in the blockchain and the real history of the data in the distributed storage practically acceptable.

File permissions in the ProvHL system are managed using the attributes of this asset containing access control lists. There are three such lists: readACL list is for access to read the file; writeACL is for access to modify the file; and execACL is for access to the file which is used either as a program for processing other files or as an input file for a data processing service. Each of these lists contains links to either users or user groups. The 'group' asset is intended just for more flexible and convenient access right management.

Virtual (overlaying) directories are also defined as assets similarly to data files and with similar corresponding set of transactions and events. In general there possible two approaches: (1) top-down approach where list of files and directories contained in a given directory is an attribute of the latter; (2) bottom-up approach where the file attribute contains the information about the directory to which this file belongs. In the ProvHL system the second approach was adopted for implementation because it essentially more simple and natural. Indeed, if to use the first approach, an operation with a file would require changing attribute values not only of the very file but also the directory which contains it. In the bottom-up approach it is enough to change only file attributes including it placement in the directory tree. To obtain the content of a directory a user sends a query to the ledger for selection of all the files with value of the directory attribute equal to ID of the directory of interest. Directories as assets can be subjected to operations similar to those in the case of files.

Data management via PMD requires a method of ensuring consensus among participants in the business process about the content and order of transactions with data.  Nowadays there exits a number of consensus algorithms that do not require resource-consuming and slow "proof-of-work" mechanism which is intrinsic for cryptocurrency blockchain networks \cite{8}.
In short, general operation of the system looks as follows: (1) users request operations with data; (2) all the requests and any other actions are recorded in the distributed ledger (that is blockchain) as provenance (in other words, chronological) metadata; (3) all the copies of the distributed ledger are identical due to rather sophisticated process of  achieving consensus between the participating parties; (4) validity of the requested transactions are checked against the politics, access control lists, smart contracts; (5) the order of the recorded transactions are defined on the basis of distributed consensus between the parties involved; (6) actions from the side of the data management service being responses to the approved user requests go through the similar validation procedure; (7) sets of valid transactions that make up the operations result in the actual data transformations in the framework of the CDCS.

\begin{figure}
\begin{center}
\includegraphics[scale=0.4]{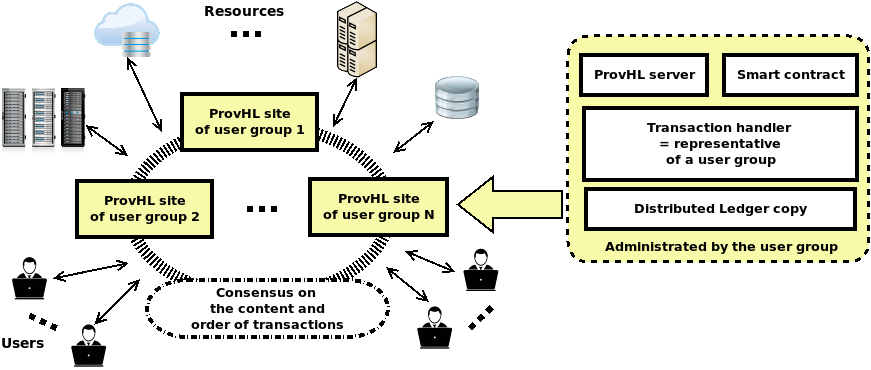}
\end{center}
\caption{Overall CDCS structure under the ProvHL management}
\label{fig:fig1}
\end{figure}

Overall CDCS structure under the ProvHL management looks like it is shown in figure~\ref{fig:fig1}.  Instead of central cervices there exists peer sites of user groups under their own administration which jointly manage the DCS according to agreed rules and politics encoded in smart contracts and consensus algorithm. 

Further details on the approach and its implementation can be found in our paper \cite{9}.

\section{Conclusion}

In many cases, participants of large projects need to combine their computing resources of into a single distributed pool. For definiteness, in this paper such a pool are called collaborative distributed computer systems (CDCS) to distinguish them from the more general concept of any distributed computer systems (DCS). To achieve combining local storages and the data in them into a single storage system one has to implement reciprocal access policies for the data of the parties involved. This implies the existence of methods for decentralized data management and access rights to them, to ensure consensus between parties involved in data sharing, on the content and procedure for operations with data. The approach proposed in this work is based on the integration of the blockchain technology, smart contracts and metadata driven data management solves these problems. A system developed on its basis, entitled ProvHL, is a completely decentralized, fault-tolerant, secure and reliable system for managing data, provenance metadata, as well as access rights to data in distributed storage. This infrastructure is free from the significant drawbacks inherent to existing solutions, namely, from the vulnerabilities associated with the presence of a central services. 

\subsection*{Acknowledgement}

This work was funded under R\&D State Assignment No. 115041410196.


\end{document}